\documentclass[JPD,10pt]{iopart}

\usepackage{amssymb}   
\usepackage{iopams}  
\usepackage{graphicx}   

\begin{document}
\title[]{Valence band structure calculations of strained Ge$_{1-x}$Sn$_x$ quantum well pFETs}
\author{H-S Lan}
\address{Graduate Institute of Electronics Engineering, Department of Electrical Engineering, National Taiwan University, Taipei 10617, Taiwan}
\ead{huangsianglan@gmail.com}
\author{C W Liu}
\address{Graduate Institute of Electronics Engineering, Department of Electrical Engineering, National Taiwan University, Taipei 10617, Taiwan}
\address{Center for Condensed Matter Sciences, National Taiwan University, Taipei 10617, Taiwan}
\address{National Nano Device Laboratories, Hsinchu 30078, Taiwan}
\ead{chee@cc.ee.ntu.edu.tw}

\vspace{10pt}
\begin{indented}
\item[]March 6 2017
\end{indented}

\begin{abstract}
The dependence of valence band structures of Ge$_{1-x}$Sn$_x$ with 0 $\leq$ $x$ $\leq$ 0.2 on Sn content, biaxial strain, and substrate orientation is calculated using the nonlocal empirical pseudopotential method. The first valence subband structure in p-type Ge cap/fully strained Ge$_{1-x}$Sn$_x$ quantum well/Ge (001) and (111) inversion layers are theoretically studied using the 6$\times$6 k$\cdot$p model. A wave-function coupling of a Ge cap with respect to a strained Ge$_{1-x}$Sn$_x$ quantum well, which is influenced by the cap thickness, valence band offset, and confined effective mass, changes the energy dispersion relation in the two-dimensional $k$-space. The increase in Sn content and the decrease in cap thickness increase the hole population in the strained Ge$_{1-x}$Sn$_x$ quantum well to reduce the transport effective mass at the zone center in the Ge/strained Ge$_{1-x}$Sn$_x$/Ge inversion layers.
\end{abstract}
%
%
%
%
\ioptwocol
\section{Introduction}

Sn incorporating into Ge potentially makes Ge$_{1-x}$Sn$_x$  alloys as a direct-gap semiconductor \cite{d1,d2,d3,d4,d5}, which make them suitable for use in n-channel field effect transistors (nFETs) owing to the small effective mass of the $\Gamma$ valley [6]. Several studies on GeSn pFETs show high hole mobilities of strained GeSn grown on Ge (001) \cite{d7, d8}, (110) \cite{d9}, and (111) \cite{d10,d11} substrates. The reduction in effective mass along the channel direction ($m_{\mathrm{chan}}$) due to biaxial compressive strain is responsible for the increased hole mobility \cite{d8, d12}. In GeSn/Ge (001), the biaxial compressive strain lifts the light hole band along the [110] direction \cite{d8}. Increasing the Sn content can further reduce the light hole effective mass \cite{d4}. In addition, the Ge cap on the GeSn channel can prevent holes from scatterings at the Ge/oxide interface \cite{d8}. Nevertheless, knowledge of valence subband structures in GeSn inversion layers is very limited owing to the unknown 6$\times$6 k$\cdot$p Luttinger parameters and deformation potential parameters of GeSn alloys. Note that linearly interpolated 6$\times$6 k$\cdot$p parameters based on the linear virtual crystal approximation (VCA) have been widely used in calculations related to valence subband structures in the SiGe inversion layers \cite{d13, d14}. Low's pioneering approach \cite{d4} fitted multiple sets of nonlinear Luttinger parameters ($\gamma_1$, $\gamma_2$, and $\gamma_3$) of relaxed GeSn alloys from full-band structures calculated using the nonlocal empirical pseudopotential method (EPM) for discrete Sn content. However, this approach increases the computational complexity, as it requires taking different strains and proportions of Sn content into account. In this work, one set of 6$\times$6 k$\cdot$p parameters for Ge$_{1-x}$Sn$_x$  alloys, where 0 $\leq$ $x$ $\leq$ 0.2 with in-plane biaxial strain of $-$3\% $\leq$ ${\varepsilon}_{||}$ $\leq$ 1\%, are fitted based on the calculated valence band structures and effective masses of strained GeSn (001), (110), and (111) wafer orientations using EPM. The fitted 6$\times$6 k$\cdot$p parameters are used to calculate the valence subband structures in inversion layers of Ge cap/10 nm GeSn quantum well (QW)/Ge (001) and (111) pFETs. Most of the holes in these cases are located in the first subband, of which favorable for carrier transport is implied by a small $m_{\mathrm{chan}}$. The dependence of $m_{\mathrm{chan}}$ of the first subband at the zone center on the Sn content, cap thickness, and wafer orientation is theoretically studied.

\section{Theoretical Framework}

\begin{figure}
\includegraphics{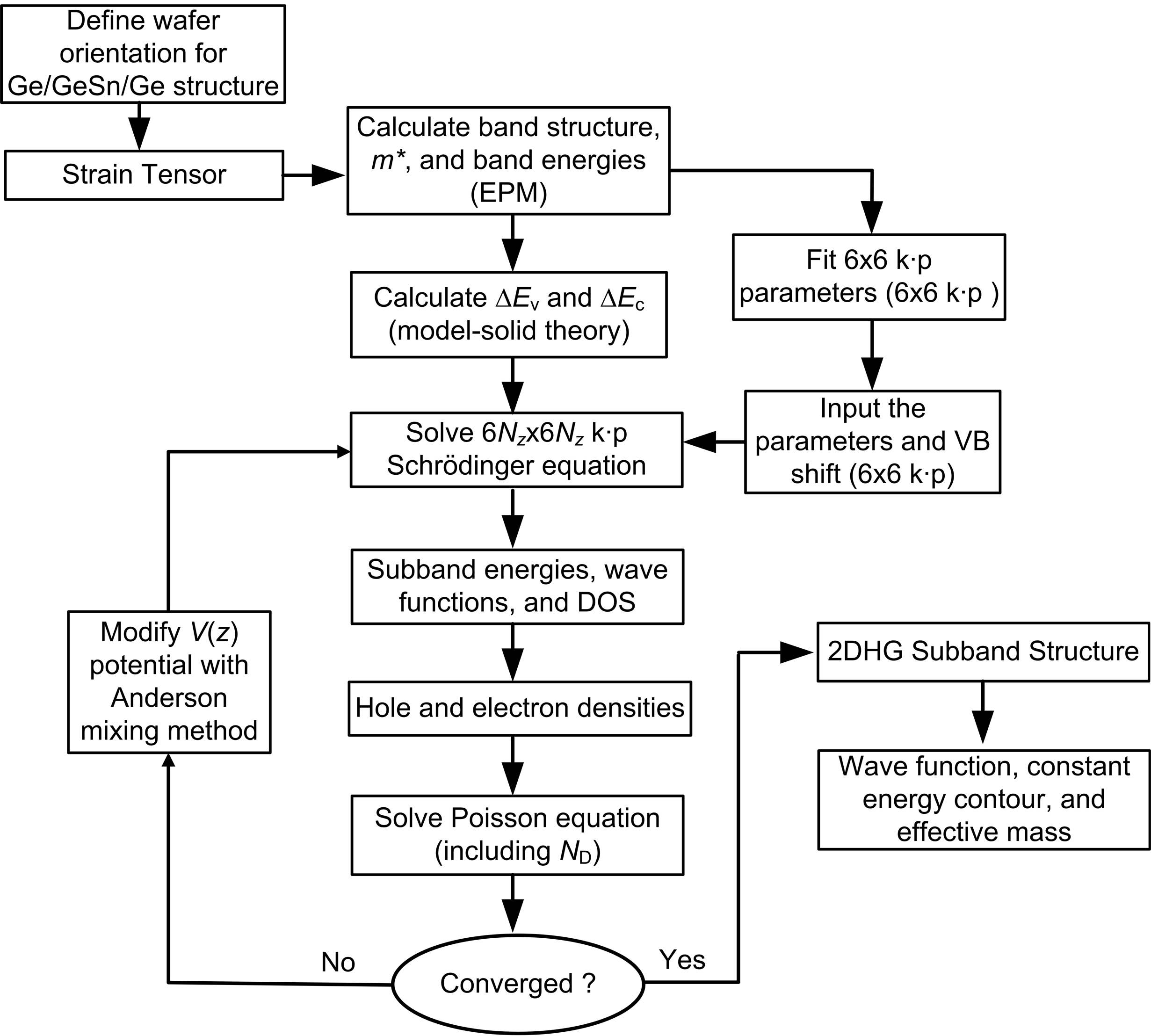} 
\caption{\label{f1}Flowchart of valence band calculations of EPM and 6$\times$6 k$\cdot$p model, and the self-consistent calculation of 6$N_z\times6N_z$ k$\cdot$p Schr\"{o}dinger and Poisson equations for the valence subband structures in the inversion layer of Ge/strained Ge$_{1-x}$Sn$_x$/Ge pFETs.}
\end{figure}

The calculation flowchart in this work is summarized in \fref{f1}. To study valence subband structures in the inversion layer of Ge cap/GeSn QW/Ge pFETs on (001) and (111) Ge substrates, the wafer orientation and the corresponding transferred strain tensor in the crystal coordinate system \cite{d15} are determined first. Effective masses (m$^{\ast}$) and valence band structures near the $\Gamma$ point ($k$ = [$-$0.05 : 0.05] in 2$\pi$/$a_0$ unit) along [100], [001], [110], [$-$110], and [111] directions, and band energies at the $\Gamma$ point of Ge$_{1-x}$Sn$_x$ with 0 $\leq$ $x$ $\leq$ 0.2 under $-$3, $-$2, $-$1, 0, and 1\% in-plane biaxial strain on (001), (110), and (111) wafer orientations are calculated using EPM and used to fit 6$\times$6 k$\cdot$p parameters through the 6$\times$6 k$\cdot$p model \cite{d13,d16}. The fitted 6$\times$6 k$\cdot$p parameters are obtained by solving nonlinear least squares with the above values calculated by EPM. Note that our EPM [17] has been reported in elsewhere and the calculated bandgaps agree with the experimental data. Linear interpolation of elastic constants ({\em C}$_{11}$, {\em C}$_{12}$, and {\em C}$_{44}$) \cite{d18}, and a bowing factor of 0.047 {\AA} \cite{d19} for the lattice constant ($a_0$) of GeSn alloys are used in the calculations.
\begin{table*}
\lineup
\caption{\label{t1}Fitted 6$\times$6 k$\cdot$p parameters.}
\footnotesize
\begin{tabular}{@{}llll}
\br
{}&Symbol&Ge&Sn\\
\mr
Luttinger parameters&$\gamma_1$&9.4415&$-$6.8605\\
{}&$\gamma_2$&2.7433&$-$5.8624\\
{}&$\gamma_3$&3.9298&$-$4.3485\\
Deformation potentials (eV)&$b$&$-$2.06&$-$2.3\\
{}&$d$&$-$7.4608&$-$4.1\\
Additional internal strain parameter (eV)&$\xi_b$&$-$0.2322&2.872\\
\br
\end{tabular}\\
\end{table*}
\normalsize

The fitted Luttinger parameters ($\gamma_1$, $\gamma_2$, and $\gamma_3$) and deformation potential parameters ($b$ and $d$) of Ge and Sn are listed in \tref{t1}. The fitted Luttinger parameters of Ge are similar with the previous reported results \cite{d20}. To take into account GeSn alloys, the 6$\times$6 k$\cdot$p parameters are expressed as a quadratic polynomial
\begin{equation}
P^{\mathrm{GeSn}}(x) = (1{-}x)P^{\mathrm{Ge}} {+} xP^{\mathrm{Sn}} {-} x(1{-}x)\theta_{P}, \label{e1}
\end{equation}
where $P$ can be $\gamma_1$, $\gamma_2$, $\gamma_3$, $b$ or $d$, and the coefficient $\theta_{P}$ is a corresponding bowing factor listed in \tref{t2}. A linearly interpolated hydrostatic deformation potential for the valence band of GeSn alloys is obtained from reference \cite{d21} ($a_{\mathrm{v}}^{Ge}$ = 2.23 eV and $a_{\mathrm{v}}^{Sn}$ = 1.58 eV). Moreover, for strained Ge$_{1-x}$Sn$_x$ (110) and (111) substrates with in-plane biaxial strain, there is a relative displacement of two atoms in the unit cell due to generated shear strains [22, 23]. An additional internal strain parameter ($\xi$) was used in the EPM calculation to take into account the internal strain effects [24]. The used $\xi$ of Ge$_{1-x}$Sn$_x$ in EPM is obtained by the linear interpolation of the reported $\xi$ of Ge \cite{d22} and Sn \cite{d25}. In 6$\times$6 k$\cdot$p model, the internal strain effect was usually ignored due to few information for parameter correction \cite{d14,d26}. To consider the effect in 6$\times$6 k$\cdot$p model, we use an additional parameter ($\xi_b$) to modify the fitted deformation potential parameter ($b$) of Ge$_{1-x}$Sn$_x$ as
\begin{equation}
b_{\mathrm{mod}}^{\mathrm{GeSn}}(x) = b^{\mathrm{GeSn}}(x) {+} (1{-}x)\xi_b^{\mathrm{Ge}} {+} x\xi_b^{\mathrm{Sn}}, \label{e2}
\end{equation}
where the $b$ of Ge$_{1-x}$Sn$_x$ is obtained from equation \eref{e1}, and the fitted $\xi_b$ of Ge and Sn are listed in \tref{t1}.

\begin{figure*}
\includegraphics{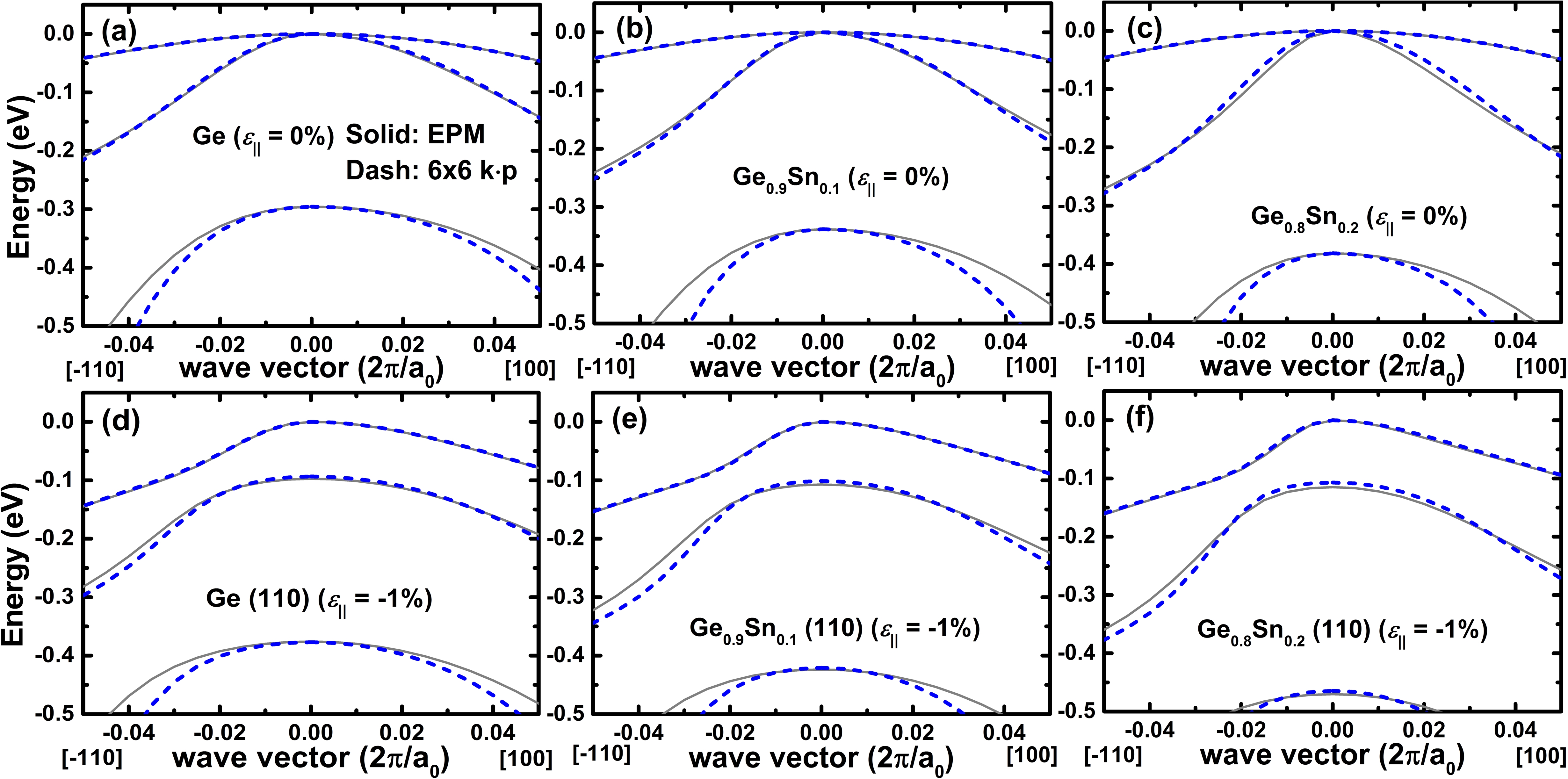} 
\caption{\label{f2}(a) The calculated valence band structures of relaxed bulk Ge, (b) relaxed Ge$_{0.9}$Sn$_{0.1}$, (c) relaxed Ge$_{0.8}$Sn$_{0.2}$, (d) strained Ge (110) with ${\varepsilon}_{||}$ = $-$1\%, (e) strained Ge$_{0.9}$Sn$_{0.1}$ (110) with ${\varepsilon}_{||}$ = $-$1\%, and (f) strained Ge$_{0.8}$Sn$_{0.2}$ (110) with ${\varepsilon}_{||}$ = $-$1\% by EPM (solid lines) and 6$\times$6 k$\cdot$p model (dashed lines).}
\end{figure*}

\begin{table}
\lineup
\caption{\label{t2}Fitted bowing factors.}
\footnotesize
\begin{tabular}{@{}llll}
\br
Bowing factor&Symbol&GeSn\\
\mr
Luttinger parameters&$\theta_{\gamma 1}$&$-$66.6075\\
{}&$\theta_{\gamma 2}$&$-$32.6386\\
{}&$\theta_{\gamma 3}$&$-$32.8118\\
Deformation potentials (eV)&$\theta_{b}$&$-$0.5502\\
{}&$\theta_{d}$&7.906\\
\br
\end{tabular}\\
\end{table}
\normalsize
The calculated valence band structures of Ge$_{1-x}$Sn$_x$ along [$-$110] and [100] directions with varying Sn content ($x$ = 0, 0.1, and 0.2) and in-plane biaxial compressive strain (${\varepsilon}_{||}$ = 0 and $-$1\%) using EPM (solid line) and 6$\times$6 k$\cdot$p (dashed line) are shown in \fref{f2}. The agreements in band energies between the two methods at the $\Gamma$ point and two uppermost band structures near the $\Gamma$ point are presented. The split-off band structure fitting is ignored to obtain best fit of the other two valence bands.

\begin{figure}
\includegraphics{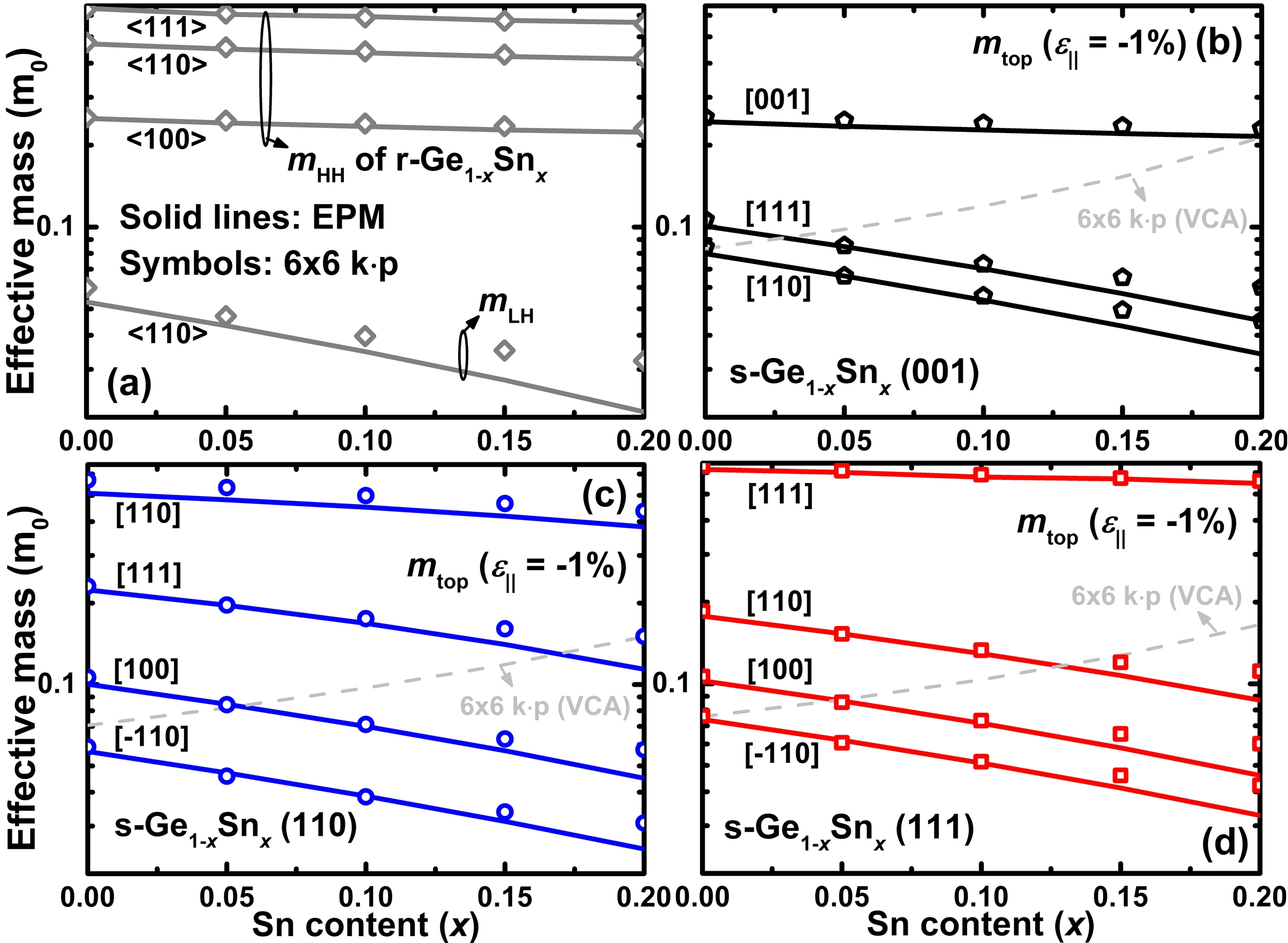} 
\caption{\label{f3}(a) The effective masses of the heavy hole band ($m_{\mathrm{HH}}$) along $\langle111\rangle$, $\langle110\rangle$, and $\langle100\rangle$ directions, and the effective mass of the light hole band ($m_{\mathrm{LH}}$) along $\langle110\rangle$ directions as a function of Sn content of relaxed Ge$_{1-x}$Sn$_x$ calculated by EPM (solid lines) and 6$\times$6 k$\cdot$p model (symbols). The calculated effective mass of the top valence band ($m_{\mathrm{top}}$) as a function of Sn content of (b) strained Ge$_{1-x}$Sn$_x$ (001) with ${\varepsilon}_{||}$ = $-$1\% along [001], [111], and [110] directions, and (c) strained Ge$_{1-x}$Sn$_x$ (110) with ${\varepsilon}_{||}$ = $-$1\% and (d) strained Ge$_{1-x}$Sn$_x$ (111) with ${\varepsilon}_{||}$ = $-$1\% along [110], [111], [100], [$-$110] directions by EPM (solid lines) and 6$\times$6 k$\cdot$p model (symbols). The calculated $m_{\mathrm{top}}$ of [110]/(001), [$-$110]/(110), and [$-$110]/(111) with ${\varepsilon}_{||}$ = $-$1\% using the reported linearly interpolated 6$\times$6 k$\cdot$p parameters \cite{d20,d27,d28} based on VCA (dashed lines) are also shown for comparison.}
\end{figure}

The calculated effective masses of the heavy hole bands ($m_{\mathrm{HH}}$) along $\langle111\rangle$, $\langle110\rangle$, and $\langle100\rangle$ directions, and light hole bands ($m_{\mathrm{LH}}$) along $\langle110\rangle$ directions of relaxed Ge$_{1-x}$Sn$_x$ (r-Ge$_{1-x}$Sn$_x$) alloys using EPM and 6$\times$6 k$\cdot$p are shown in \fref{f3}(a). The rapid decrease of $m_{\mathrm{LH}}$ along $\langle110\rangle$ direction with increasing $x$ for r-Ge$_{1-x}$Sn$_x$ is compared against the slow decrease of $m_{\mathrm{HH}}$ along $\langle111\rangle$, $\langle110\rangle$, and $\langle100\rangle$ directions. In strained Ge$_{1-x}$Sn$_x$ (s-Ge$_{1-x}$Sn$_x$) (001) with ${\varepsilon}_{||}$ = $-$1\% (\fref{f3}(b)), the effective masses along [110] and [$-$110] directions are the same due to symmetry of the band structures. For s-Ge$_{1-x}$Sn$_x$ (110) and (111) with ${\varepsilon}_{||}$ = $-$1\%, the symmetry is broken by the strain and the top valence band ($m_{\mathrm{top}}$) along [$-$110] direction is smaller than that along [110] (\fref{f3}(c) and (d)). The slow decrease of $m_{\mathrm{top}}$ along the out-of-plane direction in s-Ge$_{1-x}$Sn$_x$ (001), (110), and (111) is compared against the rapid decrease of $m_{\mathrm{top}}$ along the other directions. Note that in-plane $\langle110\rangle$ and out-of-plane directions are usually defined as channel and confined directions, respectively. The $m_{\mathrm{top}}$ calculated by 6$\times$6 k$\cdot$p based on VCA (dashed lines) along [110], [$-$110], and [$-$110] directions for s-Ge$_{1-x}$Sn$_x$ (001) (\fref{f3}(b)), (110) (\fref{f3}(c)), and (111) (\fref{f3}(d)) with ${\varepsilon}_{||}$ = $-$1\%, respectively, are also shown for comparison. The used linearly interpolated 6$\times$6 k$\cdot$p parameters are adopted from literatures \cite{d20,d27,d28}. 

After fitting 6$\times$6 k$\cdot$p parameters, the self-consistently calculation between 6$N_z\times6N_z$ k$\cdot$p Schr\"{o}dinger and Poisson equations \cite{d29,d30} is performed to calculate the two-dimensional hole gas system (2DHG) as shown in \fref{f1}. The $N_z$ denotes mesh points along the confined $z$ direction. A uniform mesh of 2 {\AA} size is used. Linear interpolation of dielectric constants of GeSn alloys are used \cite{d18}. A valence band offset ($\Delta E_{\mathrm{v}}$) at a heterointerface calculated by the model-solid theory (MST) \cite{d31,d32} are used to determine the valence band lineup potential $V(z)$ in the 6$N_z\times6N_z$ k$\cdot$p Schr\"{o}dinger equation. Note that a proposed correction term on the average valence band offset of GeSn alloys \cite{d33} is used in MST instead of the linearly interpolated average valence band offset \cite{d34}. Our reported calculated $\Delta$$E_{\mathrm{v}}$ of $\sim$80 meV in s-Ge$_{0.95}$Sn$_{0.05}$/r-Ge by MST \cite{d31} is within the reported measurement accuracy range of $\pm$ 50 meV for $x$ $\leq$ 0.08 \cite{d35}. A duplicate valence band edge (VB) shift in the $V(z)$ potential by strain in the 6$N_z\times6N_z$ k$\cdot$p matrix is removed. The treatment of the 6$N_z\times6N_z$ k$\cdot$p Hamiltonian matrix at a heterointerface is referred to reference \cite{d36}. A 2D density of state (DOS) of each subband is calculated from a tabulation of each subband structure and used to determine the hole density in the inversion layer \cite{d37}. The effective mass Schr\"{o}dinger equations are also performed to calculate the electron density in conduction band heterostructures \cite{d38}. A conduction band offset ($\Delta$$E_{\mathrm{c}}$) at a heterointerface is determined by the corresponding bandgaps and $\Delta$$E_{\mathrm{v}}$. Uniform doping density $N_{\mathrm{D}}$ of 1$\times10^{14}$ cm$^{-3}$ is used in the Poisson equation \cite{d30}. The Anderson mixing method is used to accelerate the convergence of iterative solution \cite{d39}.

\section{Results and Discussion}

\begin{figure}
\includegraphics{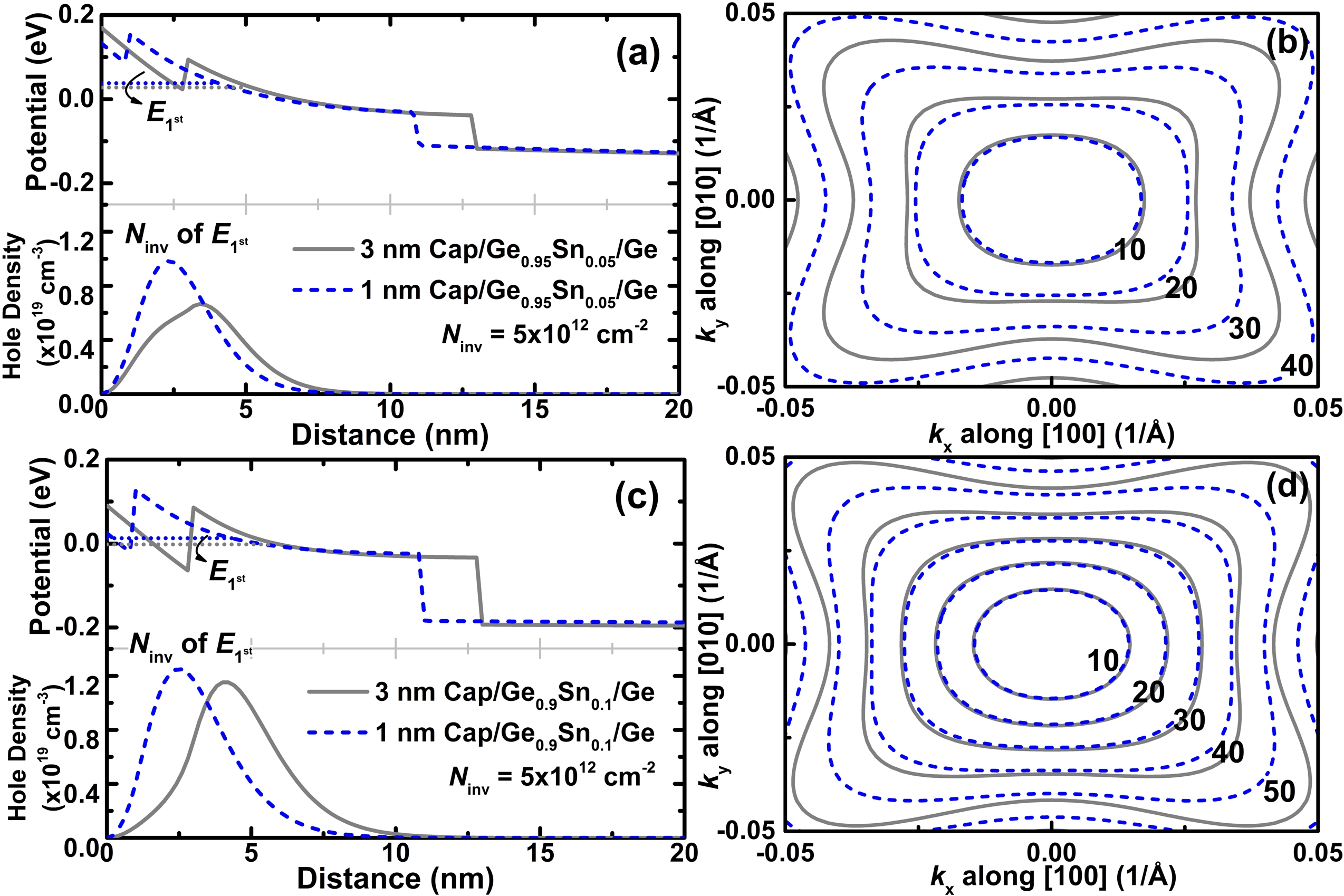} 
\caption{\label{f4}Calculated valence band diagrams of 3 nm and 1 nm Ge cap on 10 nm of (a) Ge$_{0.95}$Sn$_{0.05}$ QW and (c) Ge$_{0.9}$Sn$_{0.1}$ QW on Ge (001) substrates with the corresponding quantized energy of the first subband ($E_{1^{\mathrm{st}}}$) and first subband hole density distribution at $N_{\mathrm{inv}}$ = 5$\times10^{12}$ cm$^{-2}$. (b) and (d) Corresponding confined constant 2D energy contours of the first subbands in 3 nm (solid lines) and 1 nm cap (dashed lines) /QW/Ge (001) inversion layers. The energy interval between contours is 10 meV.}
\end{figure}

At the total inversion hole density $N_{\mathrm{inv}}$ = 5$\times10^{12}$ cm$^{-2}$, the calculated inversion layers in 3 nm and 1 nm Ge cap/10 nm s-Ge$_{1-x}$Sn$_x$ QW on Ge (001) substrate with the first subband hole density distribution are shown in \fref{f4}(a) ($x$ = 0.05) and 4(c) ($x$ = 0.1). The Fermi level energy is fixed at 0 eV in the valence band. The quantized energy of the first subband ($E_{1^{\mathrm{st}}}$) crosses the Ge cap and s-Ge$_{1-x}$Sn$_x$ QW ($x$ = 0.05 and 0.1) layers. For $x$ = 0.05 (\fref{f4}(a)), there is a substantial hole density of $E_{1^{\mathrm{st}}}$ in the 3 nm cap as compared to a low hole density of $E_{1^{\mathrm{st}}}$ in the 1 nm cap. These hole distributions influence the first subband constant 2D energy contour of the Ge/s-Ge$_{0.95}$Sn$_{0.05}$/Ge (001) inversion layer as shown in \fref{f4}(b). Note that the first subband energy dispersion can be expressed as
\begin{equation}
E(K,\phi) = E_{1^{\mathrm{st}}} {+} \frac{\hbar^2K(\phi)^2}{2m^{\ast}(K,\phi)}, \label{e3}
\end{equation}
where $K$ is the in-plane vector, $\phi$ is the polar angle in the range [0, $\pi$/2] from $k_x$, and $m^{\ast}$ is indirectly proportional to $\partial^{2}E(K,\phi)$/$\partial K(\phi)^2$ \cite{d30,d37}. The energy dispersion is hybridized by the Ge cap and s-Ge$_{0.95}$Sn$_{0.05}$ QW. The rapid change in energy (in meV) with increasing $K$ for the 1 nm cap is compared against that for the 3 nm cap (\fref{f4}(b)). For $x$ = 0.1 (\fref{f4}(c)), the hole density of $E_{1^{\mathrm{st}}}$ in the QW layer for the 3 nm and 1 nm cap are similar, which is due to the increase in $\Delta$$E_{\mathrm{v}}$ with increasing $x$. As a result, there is a similar change in energy with increase in $K$ for the first subband constant 2D energy contours of the 3 nm and 1 nm cap/s-Ge$_{0.9}$Sn$_{0.1}$/Ge (001) inversion layers (\fref{f4}(d)).

\begin{figure}
\includegraphics{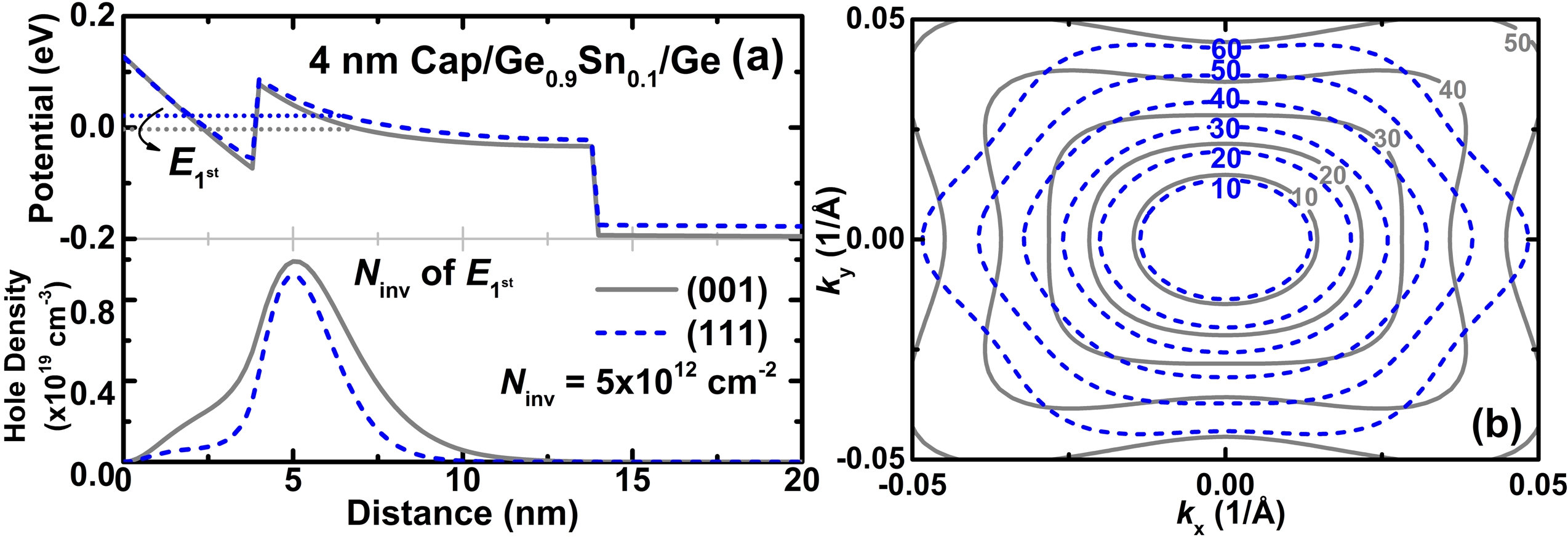} 
\caption{\label{f5}(a) Calculated valence band diagrams of 4 nm cap on 10 nm Ge$_{0.9}$Sn$_{0.1}$ QW on Ge (001) and (111) substrates with the corresponding quantized energy of the first subband ($E_{1^{\mathrm{st}}}$) and first subband hole density distribution at $N_{\mathrm{inv}}$ = 5$\times10^{12}$ cm$^{-2}$. (b) Confined constant 2D energy contours of the first subbands for the (001) (solid lines) and (111) (dashed lines) substrates. The energy interval between contours is 10 meV.}
\end{figure}

For different orientations, the calculated inversion layers of the 4 nm Ge cap/10 nm s-Ge$_{0.9}$Sn$_{0.1}$ QW on Ge (001) and (111) substrates at $N_{\mathrm{inv}}$ = 5$\times10^{12}$ cm$^{-2}$ with the corresponding first subband hole density distribution are shown in \fref{f5}(a). The hole density of $E_{1^{\mathrm{st}}}$ in the cap layer for (111) substrate is lower compared to that for (001) in spite of the similar $\Delta$$E_{\mathrm{v}}$. The $E_{1^{\mathrm{st}}}$ of (111) is more confined towards the inside than that of (001), which results in low hole density in the cap layer owing to the larger confined mass of (111) (similar to the out-of-plane values of $m_{\mathrm{top}}$ in \fref{f3}(b) and (d)). The first subband constant 2D energy contours of the Ge/s-Ge$_{0.9}$Sn$_{0.1}$/Ge (001) and (111) inversion layers are shown in \fref{f5}(b) for comparison. The $k_x/k_y$ in the device coordinate system \cite{d15} are [100]$/$[010] and [11$-$2]$/$[$-$110] for (001) and (111), respectively. To study the changes in $m_{\mathrm{chan}}$, $m^{\ast}$ at the zone center is calculated using equation \eref{e3} with $\phi$ = $\pi$/4 and $\pi$/2 for the channel direction/wafer orientation of [110]/(001) and [$-$110]/(111), respectively.

\begin{figure}
\includegraphics{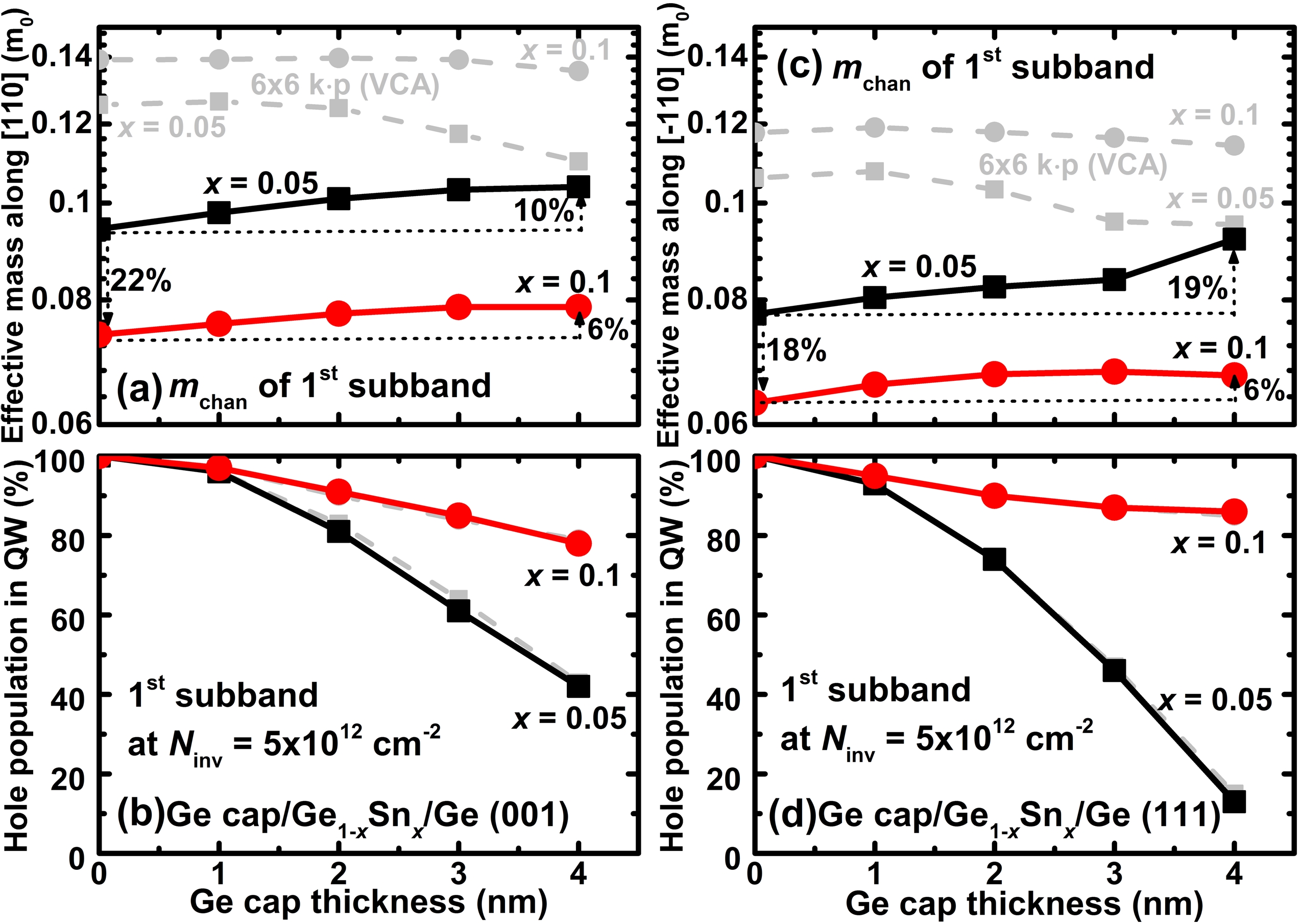} 
\caption{\label{f6}The fitted 6$\times$6 k$\cdot$p parameters and the reported linearly interpolated 6$\times$6 k$\cdot$p parameters based on VCA \cite{d20,d27,d28} are used to calculate (a) the effective mass at the zone center along [110] direction and (b) hole population in s-Ge$_{1-x}$Sn$_x$ quantum well (QW) of the first subband as a function of Ge cap thickness in the Ge/s-Ge$_{1-x}$Sn$_x$ ($x$ = 0.05 and 0.1)/Ge (001) inversion layers at $N_{\mathrm{inv}}$ = 5$\times10^{12}$ cm$^{-2}$, and (c) the effective mass at the zone center along [$-$110] direction and (d) hole population in s-Ge$_{1-x}$Sn$_x$ QW of the first subband as a function of Ge cap thickness in the Ge/s-Ge$_{1-x}$Sn$_x$ ($x$ = 0.05 and 0.1) QW/Ge (111) inversion layers at $N_{\mathrm{inv}}$ = 5$\times10^{12}$ cm$^{-2}$.}
\end{figure}

The calculated $m_{\mathrm{chan}}$ and hole population in QW as a function of cap thickness of the first subband in the Ge cap/s-Ge$_{1-x}$Sn$_x$ ($x$ = 0.05 and 0.1)/Ge (001) and (111) inversion layers at $N_{\mathrm{inv}}$ = 5$\times10^{12}$ cm$^{-2}$ are shown in \fref{f6}. The $m_{\mathrm{chan}}$ in the (111) inversion layer (\fref{f6}(c)) is smaller than that in the (001) inversion layer (\fref{f6}(a)) owing to quantum confinement. This is different from the trend seen in bulk effective masses of s-Ge$_{1-x}$Sn$_x$ [110]/(001) (\fref{f3}(b)) and [$-$110]/(111) (\fref{f3}(d)). Note that the quantum confinement also causes drastic warped valence subband structures in the case of (110) \cite{d40,d41}, which is not discussed here. The $m_{\mathrm{chan}}$ is reduced with increasing Sn content or decreasing cap thickness. This trend is contrary to the increased $m_{\mathrm{chan}}$ calculated by 6$\times$6 k$\cdot$p (VCA). In the case of 0 nm cap, there is $\sim$20\% reduction of $m_{\mathrm{chan}}$ with an increase of 5\% in Sn. The increase in $m_{\mathrm{chan}}$ with increasing cap thickness is approximately similar to the increase in the average of $m_{\mathrm{chan}}$ in Ge cap and s-Ge$_{1-x}$Sn$_x$ QW weighted by its corresponding hole population.


\section{Summary}
One set of 6$\times$6 k$\cdot$p Luttinger and deformation potential parameters for valence band structures and effective masses of Ge$_{1-x}$Sn$_x$ (001), (110), and (111) wafer orientations, where 0 $\leq$ $x$ $\leq$ 0.2 with in-plane biaxial strain of $-$3\% $\leq$ ${\varepsilon}_{||}$ $\leq$ 1\% are fitted based on the calculated valence band structures by EPM. The internal strain effects by shear strain were taken into account in the EPM and 6$\times$6 k$\cdot$p calculations. The first valence subband structures in the p-type Ge cap/10 nm s-Ge$_{1-x}$Sn$_x$ QW/Ge (001) and (111) inversion layers were theoretically studied using the self-consistent calculation of 6$N_z\times6N_z$ k$\cdot$p Schr\"{o}dinger and Poisson equations. The valence band lineups at the heterointerfaces were theoretically determined using the model-solid theory. In the QW structures, the (111) has a lower $m_{\mathrm{chan}}$ in the first subband than (001) owing to quantum confinement. The Sn content controls the effective mass and $\Delta$$E_{\mathrm{v}}$, and the $\Delta$$E_{\mathrm{v}}$ and cap thickness control the hole population in the QW to affect the $m_{\mathrm{chan}}$ of the first subband. Using the fitted 6$\times$6 k$\cdot$p parameters, the increase in Sn content and the decrease in cap thickness reduce the $m_{\mathrm{chan}}$ as compared to the incorrect increased $m_{\mathrm{chan}}$ using the reported linearly interpolated 6$\times$6 k$\cdot$p parameters of Ge and Sn.
\ack This work was supported by Ministry of Science and Technology, Taiwan, R.O.C under Grant Nos. 105-2622-8-002-001-, 105-2911-I-009-301, and 103-2221-E-002-232-MY3. The support of high-performance computing facilities by the Computer and Information Networking Center, National Taiwan University, is also highly appreciated.

\end{document}